\documentclass{vgtc}                          

\ifpdf
  \pdfoutput=1\relax                   
  \usepackage{graphicx}                
  \DeclareGraphicsExtensions{.pdf,.png,.jpg,.jpeg} 
\else
  \usepackage{graphicx}                
  \DeclareGraphicsExtensions{.eps}     
\fi%

\graphicspath{{figures/}{pictures/}{images/}{./}} 

\usepackage{microtype}                 
\PassOptionsToPackage{warn}{textcomp}  
\usepackage{textcomp}                  
\usepackage{mathptmx}                  
\usepackage{times}                     
\usepackage{cite}                      
\usepackage{tabu}                      
\usepackage{booktabs}                  

\onlineid{0}

\vgtccategory{Research}

\vgtcinsertpkg

\title{City Planning with Augmented Reality} 

\author{Catherine Angelini\thanks{e-mail:c.angelini10@gmail.com}\\ %
        \scriptsize Florida International University %
\and Adam S. Williams\thanks{e-mail:AdamWil@csu.edu}\\ %
     \scriptsize Colorado State University	 %
\and Mathew Kress\thanks{e-mail:mkres006@fiu.edu}\\ %
     \scriptsize Florida International University
\and Edgar Ramos Vieira\thanks{e-mail:evieira@fiu.edu}\\ %
     \scriptsize Florida International University
\and Newton D'Souza\thanks{e-mail:ndsouza@fiu.edu}\\ %
     \scriptsize Florida International University
\and Naphtali D. Rishe\thanks{e-mail:NDR@acm.org}\\ %
     \scriptsize Florida International University
\and Joseph Medina\thanks{e-mail:jmedi145@fiu.edu}\\ %
     \scriptsize Florida International University
 \and Francisco Ortega\thanks{e-mail:fortega@colostate.edu}\\ %
     \scriptsize Colorado State University	
}

\abstract{We present an early study designed to analyze how city planning and the health of senior citizens can benefit from the use of augmented reality (AR) using Microsoft's HoloLens. We also explore whether AR and VR can be used to help city planners receive real-time feedback from citizens, such as the elderly, on virtual plans, allowing for informed decisions to be made before any construction begins.}

\CCScatlist{
  \CCScatTwelve{Human-centered computing}{Visu\-al\-iza\-tion}{Visu\-al\-iza\-tion techniques}{Treemaps};
  \CCScatTwelve{Human-centered computing}{Visu\-al\-iza\-tion}{Visualization design and evaluation methods}{}
}

\begin{document}


\firstsection{Introduction}

\maketitle

We present an early study designed to analyze how city planning will benefit from the use of AR and VR, making it more efficient and accurate in catering to the needs of people such as senior citizens. In the study, AR and VR are used to help understand what elements would make the participants more willing to exercise in a specific area of Miami. This area is referred to as the ``M-path,'' which is an underutilized zone in Miami, FL, that lies under Miami's elevated  Metro-rail transit system and passes through several inner-city neighborhoods, where a large population of senior citizens is located and is known for its heavy vehicular traffic. This study introduces a novel way to address this issue by introducing a groundbreaking technology to enhance a city's planning efforts. 
\section{Related Work}


Due to how recently the AR technology has been adopted, limited research has supported city viewing or planning in this way. For example, Zhang et al. used the Microsoft HoloLens to visualize the City of Toronto, Canada~\cite{Zhang2018}. Even though the visualization was of a miniature model on top of a table, it shed light on the potential of applying AR to city planning as entire buildings, roads, parks and more could be visualized and modified before construction begins. On the other hand, Claes and Moore used AR to increase awareness about local issues around a city~\cite{Claes2013}.  There have also been studies to improve health and safety using AR. For example, HybridPlay used AR technology to increase outside physical activity \cite{Boj2018-bk}. With this application, users could attach a sensor to playground equipment to be able to use the equipment as an input to their phone to complete mini-games. They found that the use of such games increases the likelihood of users going to a park or outside to exercise.

\section{System Design}

In order to understand what future elements would take for senior citizens to exercise in this area, we decided to analyze how AR could be used for them to see the changes superimposed directly on the site, using the Microsoft HoloLens headset and Unity Game Engine to create the environment, and give real-time feedback on those changes. The virtual environment was based upon actual measurements collected on site and the problem areas that were identified by the city and researchers, which included the following: lack of places to sit, lack of safety, noise from the highway, no clear path for walking or cycling, darkness at night, no nearby places to use the restroom, and the overall unsafe atmosphere of the area. 

\begin{figure}[ht!]
 \centering 
 \includegraphics[width=\columnwidth]{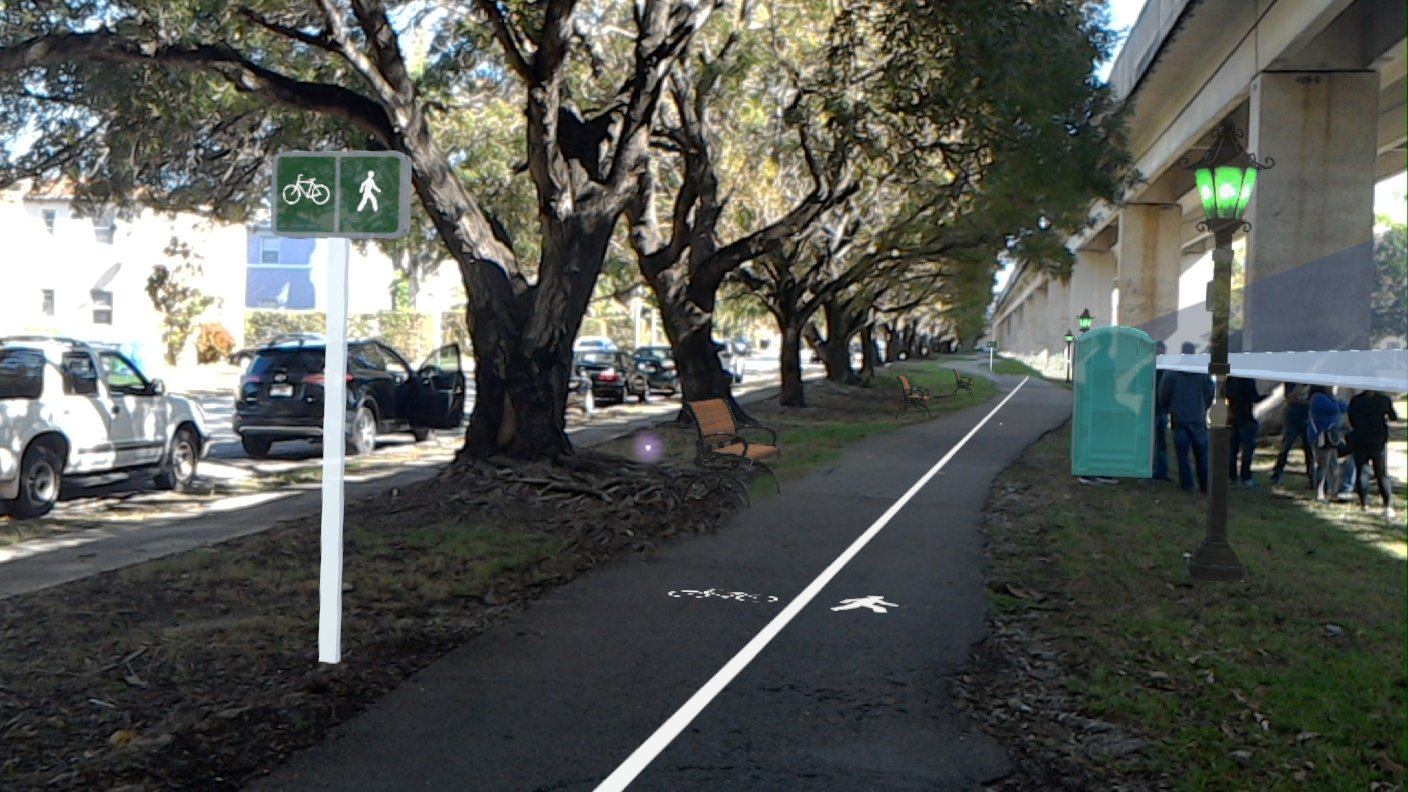}
 \caption{Environment in AR showing the glass wall scene}
 \label{fig:environmentar}
\end{figure}

As a result, we decided to focus on receiving feedback on three separate scenes, each including lamp posts, separate pedestrian and cyclist lanes, signage/way finding, either a small wall, a tall wall, or a glass wall to separate and minimize highway noise, benches, and bathroom pods. However, we also received positive feedback regarding the HoloLens itself. All of the distances and positioning of the elements relative to each other were accurate to the measurements we took. Objects remained fixed in their spatial location as the participants moved around and walked by them. The final environment can be seen in Figure \ref{fig:environmentar}. Note that the wall seems to be floating above the ground. The HoloLens image capture uses a different perspective than the headset's wearer sees. To the participants the objects appeared in their correct locations. Hence, it was evident that the use of AR allowed the participants to walk through, see and experience the site with all the virtual modifications, giving them a detailed look at how the environment could look in the future.  



\section{Experiment Design}

For this study, 10 elderly adults who live within 2 miles of the ``M-path'' near the [City] Metro-rail Station participated. The participants were required to be at least 60 years old, be English or Spanish speakers, be able to walk independently without any walking aids, have no lower limb surgery or injuries from falls during the previous 6 months, and pass the Mini-Cog test\footnote{See \url{https://mini-cog.com/}.}. All participants received an explanation about the study's goals and procedures and were given the opportunity to ask and receive answers to any questions they had. After the explanations, they were asked to read and sign the informed consent form in their preferred language and were given \$30 for their participation.

\subsection{Preparation}

The participants were fitted with Inertial Measurement Units (IMUs), from the MTw Awinda system from Xsens Technologies, to assess their gait and balance during testing. This was done to ensure the AR or VR headsets would not have a negative impact on their mobility. During testing, the participants walked along the existing path on the site with the equipment. A control experiment was also conducted as a simple walk-through without the headsets. The first part of the experiment involved assessing their balance while standing still and getting accustomed to the AR and VR experience. Hence, after confirming that the IMUs were not disrupting the participants mobility, they were asked to stand on a force plate for 30 seconds to measure their balance, cross-referencing the force plate with the MTw Awinda system for calibration. The HoloLens was then placed on the subject powered off and their balance was measured for 30 seconds. Then the HoloLens was turned on and prepared for the participant with the scene with the tall wall and their balance was taken again. Then an HTC Vive was used with the participant standing to allow them to first fully see the entire virtual environment with no external distractions. This way they would know what new elements were going to appear on the AR environment mixed in with the real world.


\subsection{Walk-through}

After the previous steps mentioned above were completed, the participant walked four times in the given path using the HoloLens headset with someone following right behind them for safety purposes. It was of particular importance to maintain a natural walking condition while having someone close to the subjects in case they fell due to loss of balance. The experiment began with the HoloLens placed on participant's head while turned off, followed by it being turned on and including the short wall, tall wall, and glass wall scenes, with each condition being randomized. 

\subsection{Participant Response}

At the end of the experiment, the participants were asked to rate how much the street furniture contributed to the environment on a Likert scale (1-5). Most responses about the virtual objects were favorable. The results should be taken as a simple guide., but it is noteworthy that to share that the participants felt that those objects were well placed for this particular part of the city.  For instance, when asked which wall they preferred, 18.2\% wanted the small wall, 27.3\% liked the tall wall, and 54.5\% liked the glass wall. The average likert responses (0-5) to how much people enjoyed different aspects of the environment were as follows: lamp post 4.7, Cycle lane 4.8, signs 4.1, benches 4.7, bathrooms 4.625, small wall 3.6, tall wall 3.3, glass wall 4.4.



\section{Discussion}

AR made a difference in their understanding of how the area could look in the near future. Several of the participants had attended previous city meetings where there were talks about the ``M-path'' and pictures were shown of how it could potentially look. However, the subjects still had questioned whether themselves or other people in their community would actually use the park. Informally, we were told by the participants that it was not until they saw and experienced the actual virtual environment while using the HoloLens that they were able to internalize the changes proposed by the city. Implying that being able to personally experience a virtual design superimposed onto a real location would increase the quality of future designs due to the real-time feedback city planners can receive. 

The participants felt that the holograms that they saw around them of the benches, lampposts, and other street furniture, immersed them into a world where the unsafe and isolated park they were familiar with was transformed into a safe, modern, and elderly-friendly park that motivated them to be go outside and be active.

\section{Design Implications}

We analyzed how AR and VR can work outside to re-imagine areas by blending the real and virtual world together to create an immersive and realistic experience for users. We also examined how AR and VR can make an impact in the architecture and health. There is an up and coming reality where the construction of a home for a family could take place with a virtual model being shown to them through AR to gauge their response. They could walk through the house in scale and make decisions on each element before any of it has even been constructed. d. Similarly, city planners and architects will be able to build shopping centers, buildings, and anything that they can think of.  

After conducting this study we have some recommendations on designing these types of virtual spaces. The 3D space should be built with bright and light colors otherwise they will appear less transparent when rendering. Due to the HoloLens' low field of view of 35 degrees in the current version, the use of the Vive VR headset, with its larger field of view of 110 degrees allowed the users to see the entire area without having to move their head or walk around. Another important consideration is to try to rely on applications that don't need internet connectivity.

The implications that this research may have for city planning and health promotion have direct impacts in architecture, construction, and other areas. There are cases, such as the one presented in this work, where constructing different options is either unfeasible or economically counterproductive. We have shown how virtual environments such as those made possible through AR headsets can have an impact in making these decisions easier. AR can make city planning extremely intuitive and integrated with the general public, especially in an area as critical as our elderly and their health. 

\section{Future Work and Conclusion}


After taking into consideration the participants' requests or concerns for street furniture outside of what they had seen on the AR and VR headsets and their choice of wall, we will share our results with the City of Miami and add the suggested models to the applications to create a final AR environment with the participants' choices included. This design will be used to build the actual ``M-path'' park in 2019. Ideally, we are planning to conduct a larger experiment with over 30 participants with the new recommendations.

While more research is needed, there are indications that this technology can provide a real benefit to accessible city planning and improve the health of senior citizens. The participants expressed great satisfaction and enthusiasm with respect to wearing the AR headset and how immersive the experience felt for them. We have also shown that this technology can be safely and easily used by even senior citizens, proving it to be an accessible and powerful technology. With this technology, cities can cater to the needs of their citizens by being able to see and analyze in advance when planning for smarter, better and more accessible environments.

\bibliographystyle{abbrv-doi}

\bibliography{cityproject}
\end{document}